\documentclass[pre,showpacs,preprint]{revtex4}

\usepackage{amsmath}

\usepackage{graphicx}
\usepackage{dcolumn}
\usepackage{bm}

\begin{document}

\title{The shearing instability of a dilute granular mixture}
\author{J. Javier Brey and M.J. Ruiz-Montero}
\affiliation{F\'{\i}sica Te\'{o}rica, Universidad de Sevilla,
Apartado de Correos 1065, E-41080, Sevilla, Spain}
\date{\today }

\begin{abstract}
The shearing instability of a dilute granular mixture composed of smooth inelastic hard spheres or disks is investigated.
By using the Navier-Stokes hydrodynamic equations, it is shown that the scaled transversal velocity mode exhibits a divergent behaviour, similarly to what happens in one-component systems. The theoretical prediction for the critical size is compared with direct Monte Carlo simulations of the Boltzmann equations describing the system, and a good agreement is found. The total energy fluctuations in the vicinity of the transition are shown to scale with the second moment of the distribution. The scaling distribution function is the same as found in other equilibrium and non-equilibrium phase transitions, suggesting the existence of some kind of universality.

\end{abstract}


\maketitle

\section{Introduction}
\label{s1}
The simplest statistical mechanics model for granular gases is a system of smooth inelastic hard spheres or disks, with constant coefficient of normal restitution \cite{Go03,BDyS97}. One characteristic feature of these systems, as compared with their elastic limit, is their instability against small wave-vector spatial fluctuations when they evolve freely \cite{GyZ93,MnyY94,ByP04}.  In the case of one-component granular gases, this spontaneous symmetry breaking leads to the formation of velocity vortices and density clusters, being often referred to as the shearing or clustering instability of the homogeneous cooling state (HCS). It is accurately predicted by a linear stability analysis of the hydrodynamic Navier-Stokes equations of granular gases, which also shows that it is driven by the transversal shear mode \cite{GyZ93,BRyC99}. Moreover, fluctuating hydrodynamics is able to predict not only the initial set-up of the spatial correlations \cite{vNEByO97,BMyR98}, but also the behaviour of the system near the critical point of the instability \cite{BDGyM06}. This includes the critical exponents governing the behaviour of both macroscopic properties and fluctuations.

In the last years, the hydrodynamic theory of granular gases has been extended to binary mixtures. Navier-Stokes equations for the hydrodynamic fields of the mixture, with explicit expressions for the involved transport coefficients, have been derived \cite{GyM07,GHyD07}. The hydrodynamic equations for a mixture are much more involved than those for a one-component system and, therefore, so is the hydrodynamic linear stability analysis of the HCS \cite{GMyD06}.  But there is no reason to expect that the physical mechanisms leading to the shearing instability in simple granular gases does not hold for mixtures. If that is the case, the behavior of the transversal component of the velocity field as the size of the system increases is the origin of the instability. And it happens that the evolution equation for this hydrodynamic mode is decoupled from the equations for the rest. This feature greatly simplifies the analysis of the initial set up of the instability.

The aim of this paper is to investigate the behavior of the transversal velocity mode in a binary granular mixture identifying, in particular, the existence of the shearing instability and determining the critical point predicted by the hydrodynamic theory. Also the behaviour of some average properties of the granular mixture in the vicinity of the instability will be  studied and compared with those of a single component granular gas. In addition, the critical behaviour of some quantities that are peculiar of granular mixtures, as the non-equipartition of kinetic theory will be investigated. It is worth to emphasize that both theory and simulations presented here are restricted to the linear regime in which deviations of the fields from their values in the HCS are small.

The remaining of this paper is organized as follows. In Sec. \ref{s2}, the relevant properties of the HCS of a granular mixture in the context of kinetic theory are summarized. This includes the criterion determining the partial temperatures of both components of the mixture. Section \ref{s3} consists of a short review of the linear stability analysis of the transversal velocity field of the HCS to Navier-Stokes order \cite{GMyD06}. The associated eigenvalue is identified, and it is shown that it has a qualitative change of behaviour when the size of the system is larger than a critical value, which  depends on the parameters defining the system. In Sec.\ \ref{s4}, the dynamics of the inelastic hard spheres or disks is reformulated by means of a change of variable, so that the HCS is mapped onto a steady state. It is a straightforward  extension of a method previously developed for one-component granular systems. This steady representation is used in Sec.\ \ref{s5} to perform direct Monte Carlo simulations (DSMC) of the system, whose results are compared with the theoretical predictions. The behaviour of the total energy fluctuations is also analyzed using the simulation results. It is seen that the relative dispersion of the energy fluctuations exhibits a power-law divergent behaviour near the instability. The paper concludes with a short discussion of the results and an analysis of the shape of the probability density distribution for the total energy fluctuations.

\section{The homogeneous cooling state of a dilute granular mixture}
\label{s2}
A fluidized binary mixture of smooth inelastic hard spheres ($d=3$) or disks ($d=2$) is considered. The mass and diameter of particles of species $i$ ($i=1,2$) are $m_{i}$ and $\sigma_{i}$, respectively. The inelasticity of collisions is assumed to be described by constant, velocity independent, coefficients of normal restitution. There are three of them: $\alpha_{11}$, $\alpha_{22}$, and $\alpha_{12}= \alpha_{21}$, where $\alpha_{ij}$ refers to the collision of a particle of species $i$ and a particle of species $j$. These coefficients are defined in the interval $0 < \alpha_{ij} \leq 1$, the value unity being the limit of elastic collisions.

The macroscopic fields number densities $n_{i}({\bm r},t)$, flow velocity ${\bm u}({\bm r},t)$, and granular temperature $T({\bm r},t)$ are defined in the usual way as local velocity moments of the distribution function of the system, although setting the Boltzmann constant equal to unity. More precisely, they can be expressed in terms of the one-particle distribution functions of the two species $f_{i}({\bm r}, {\bm v},t)$ as
\begin{equation}
\label{2.1}
n_{i}(\bm{r},t)=\int d{\bm v}\, f_{i}({\bm r},{\bm v},t),
\end{equation}
\begin{equation}
\label{2.2}
\rho ({\bm r},t)  {\bm u}(\bm{r},t)= \sum_{i=1,2} \int d{\bm v}\, m_{i} {\bm v} f_{i}({\bm r},{\bm v},t),
\end{equation}
\begin{equation}
\label{2.3} n(\bm{r},t) T(\bm{r},t)= \sum_{i=1,2} \int d{\bm v}\, \frac{ m_{i}V^{2}({\bm
r},t)}{d} f_{i}({\bm r},{\bm v},t),
\end{equation}
where $\rho \equiv m_{1}n_{1}+m_{2}n_{2}$ is the total mass density, $n \equiv n_{1}+n_{2}$ is the total number density, and ${\bm V} \equiv {\bm v}- {\bm u}$ is the peculiar velocity.

In this paper, attention will be restricted to a low density system. Then, the time evolution of the one-particle distribution functions is given by a pair of coupled nonlinear Boltzmann equations \cite{BDyS97}
\begin{equation}
\label{2.4}
\left( \partial_{t} +{\bm v} \cdot {\bm \nabla} \right) f_{i} ({\bm r},{\bm v},t) = \sum_{j=1,2} J_{ij} \left[{\bm r},{\bm v},t|f_{i},f_{j} \right],
\end{equation}
$i=1,2$, and $J_{ij}$ denoting the Boltzmann collision operator describing the scattering of pairs of particles $i,j$. From Eqs.\ (\ref{2.4}), balance equations for the macroscopic fields are derived by multiplying by $1$, ${\bm v}$, and $v^{2}$, respectively, and subsequent integration over the velocity. They have the form
\begin{equation}
\label{2.5}
\partial_{t}n_{i}+\nabla \cdot \left( n {\bm u}+
 {\bm j}_{i} \right)=0,
\end{equation}
\begin{equation}
\label{2.6}
\partial_{t} {\bm u}+{\bm u} \cdot \nabla {\bm u}+ \rho^{-1} \nabla \cdot
{\sf P} =0,
\end{equation}
\begin{equation}
\label{2.7}
\partial_{t} T  +  {\bm u} \cdot \nabla T - \frac{T}{n} {\bm \nabla} \cdot \sum_{i} {\bm j}_{i}
 + \frac{2}{nd} \left( \nabla \cdot {\bm q} + {\sf P} : \nabla
{\bm u} \right) +T \zeta=0\, .
\end{equation}

In the above expressions, ${\bm j}_{i}$ is the number of particles flux for species $i$ relative to the local flow, ${\sf P}$ is the pressure tensor, ${\bm q}$ is the total heat flux, and $\zeta$ is the cooling rate giving account of the loss of energy in collisions. These quantities are defined as functionals of the one-particle distribution functions.

The balance equations (\ref{2.5})-(\ref{2.7}) admit time-dependent homogeneous solutions characterized by uniform densities $n_{i,h}$, a vanishing velocity field ${\bm u}_{h}={\bm 0}$, and an homogeneous granular temperature $T_{h}$  evolving in time accordingly with
\begin{equation}
\label{2.8}
\partial_{t} T_{h} (t) = - \zeta_{h}(t) T_{h}(t),
\end{equation}
where $\zeta_{h}$ is the cooling rate of the homogeneous state. Of course, this equation is only meaningful if the cooling rate is expressed in terms of $n_{h}$ and $T_{h}$, then becoming a closed equation for $T_{h}$. This is accomplished when the distribution functions of both species depend on time only through the granular temperature $T_{h}(t)$. The distribution functions $f_{i,h}({\bm v},t)$ having this property are said to be ``normal                             '' and define the homogeneous cooling state (HCS) of the mixture \cite{GyD99}, which is the state considered in this paper.

Partial temperatures of the species in the HCS, $T_{i,h}(t)$, are defined through
\begin{equation}
\label{2.9}
n_{i,h} T_{i,h}(t) = \int d{\bm v}\, \frac{m_{i}v^{2}}{d} f_{i,h}({\bm v},t).
\end{equation}
Therefore, it is
\begin{equation}
\label{2.10}
\sum_{i=1,2} n_{i,h} T_{i,h} (t) =n_{h} T_{h}(t).
\end{equation}
Evolution equations for the partial temperatures are directly derived from the Boltzmann equations, particularized for the HCS,
\begin{equation}
\label{2.11}
\partial_{t} T_{i,h}(t) = - \zeta_{i,h}(t) T_{i,h}(t),
\end{equation}
with the partial cooling rates $\zeta_{i,h} (t)$  given by
\begin{equation}
\label{2.12}
\zeta_{i,h}(t)=-\frac{1}{n_{i,h} T_{i,h}(t)d} \sum_{j} \int d{\bm v}\, m_{i} v^{2}
J_{i,j}[{\bm v}|f_{i,h},f_{j,h}] \, ,
\end{equation}
Consistency of Eqs.\ (\ref{2.8}), (\ref{2.10}), and (\ref{2.11}) requires that
\begin{equation}
\label{2.13}
n_{h} T_{h}(t) \zeta_{h}(t) = \sum_{i=1,2} n_{i,h} T_{i,h}(t)\zeta_{i,h}(t).
\end{equation}
Moreover, as a consequence of the distribution functions of the HCS being normal it is \cite{GyD99}
\begin{equation}
\label{2.14}
\zeta_{1,h}(t)=\zeta_{2,h}(t)=\zeta_{h}(t).
\end{equation}

The explicit evaluation of the cooling rates requires us to solve the coupled pair of Boltzmann equations for the distribution functions of the HCS. Nevertheless, a quite accurate approximation, at least for not very strong inelasticities, is obtained by using Gaussian distributions for $f_{i,h}({\bm v},t)$ with the second moments corresponding to the partial temperature $T_{i,h}(t)$. In this way, it is obtained \cite{GyD99,DByL02,ByT02,GyM07},
\begin{equation}
\label{2.15}
\zeta_{i,h}=\frac{4\pi^{(d-1)/2}}{\Gamma\left(\frac{d}{2}\right) d} v_{0}(T_{h}) \lambda_{h} \sum_{j=1,2} x_{j} \mu_{ji} \left(\frac{\theta_{i}
 +\theta_{j}}{\theta_{i}\theta_{j}}\right)^{1/2} (1+\alpha_{ij})\left[1-\mu_{ji}\frac{1+\alpha_{ij}}{2}
 \frac{(\theta_{i}+\theta_{j}}{\theta_{j}}\right] \left( \frac{\sigma_{ij}}{\sigma_{12}} \right)^{d-1}   ,
 \end{equation}
where $x_{i} \equiv n_{i}/n$ is the number concentration of species $i$, $\sigma_{ij} \equiv (\sigma_{i}+ \sigma_{j})/2$,
\begin{equation}
\label{2.16}
v_{0}(T) \equiv \left( \frac{2T}{\mu} \right)^{1/2},
\end{equation}
with
 \begin{equation}
 \label{2.16a}
\mu \equiv \frac{m_{1} m_{2}}{m_{1}+m_{2}},
\end{equation}
is a thermal velocity,
\begin{equation}
\label{2.17}
\lambda_{h} \equiv \left( n_{h} \sigma_{12}^{d-1} \right)^{1/2}
\end{equation}
is a characteristic length,
\begin{equation}
\label{2.18}
 \mu_{ij} \equiv \frac{m_{i}}{m_{i}+m_{j}}\, ,
 \end{equation}
 and
 \begin{equation}
 \label{2.19}
 \theta_{i} \equiv  \frac{m_{i} T_{h}}{\mu T_{i,h}}.
\end{equation}
Now, the  expressions of $\zeta_{1,h}$ and $\zeta_{2,h}$ can be introduced into Eq. (\ref{2.14}). The solution of the resulting equation provides $\gamma \equiv T_{1,h}(t) /T_{2,h}(t)$ as a function of $n_{1,h}/n_{2,h}$ and the other parameters defining the model \cite{GyD99,ByT02,GyM07,ByR11}.
The accuracy of the results derived in this way has been checked by comparing with molecular dynamics results \cite{DHGyD02,MyG02,RyB13a}, as well as with data obtained from particle simulations of the Boltzmann equations using the DSMC method.

\section{Linear analysis of the transversal shear mode of a granular mixture}
\label{s3}
In the case of a one-component granular system, it is well known that the HCS becomes unstable when a linear size of the system is larger than the critical value, which depends on the parameters defining the state \cite{GyZ93,MnyY94}. For dilute granular gases, a linear stability analysis of the hydrodynamic Navier-Stokes equations shows that the origin of the instability lies in the behaviour of the transversal velocity mode. For sizes of the system larger than the critical one, the transversal velocity decays slower than the square root of the temperature \cite{BDKyS98}.  As a consequence, nonlinear coupling of the scaled modes becomes relevant and a clustering instability develops in the system. It is worth to stress that the transversal component of the velocity field is not itself linearly unstable, but it enslaves the other hydrodynamic modes through some nonlinear coupling. In particular, this leads to an increase of the temperature in the regions of larger vorticity. Then, a pressure gradient shows up and produces a density fluctuation leading to the formation of clusters. A detailed account of the theory and the comparison with simulation results is given in \cite{BRyC99}.

For a binary granular mixture, the complexity of the Navier-Stokes hydrodynamic equations \cite{GyM07,GHyD07} makes it more complex the full linear stability analysis of the HCS. Nevertheless, in the mixture the scaled transversal velocity mode is decoupled from the other hydrodynamic modes, as it happens for one-component granular gases \cite{GMyD06}. Consequently, the analysis of the transversal mode is quite simple. Here the change in its time behaviour will be investigated. This change indicates the need of considering nonlinear couplings between modes and can be considered as the precursor of a clustering instability, by analogy with the one-component case.

Consider the evolution equation for the velocity, derived from the momentum conservation, Eq. (\ref{2.6}). To Navier-Stokes order, the pressure tensor ${\sf P}$ has been computed by using the Chapman-Enskog method to solve the pair of coupled Boltzmann equations of the mixture. It reads \cite{GyD02,GyM07}
\begin{equation}
\label{3.1}
{\sf P} =p {\sf I}-\eta \left[ (  \nabla {\bm u})+(  \nabla {\bm u})^{+}-\frac{2}{d} (  \nabla \cdot {\bm u}) {\sf I}\right] \, .
\end{equation}
In this expression, $p=nT$ is the local pressure, ${\sf I}$ is the unit tensor of dimension $d$, and $\eta$ is the coefficient of shear viscosity of the mixture, that can be written as
\begin{equation}
\label{3.2}
\eta = \frac{p \lambda}{v_{0}(T)}\, \eta^{*},
\end{equation}
where $\lambda \equiv (n \sigma_{12}^{d-1} )^{-1}$ and $v_{0}(T)$ has been defined in Eq. (\ref{2.16}).  Moreover, $\eta^{*}$ is a dimensionless function of the partial temperatures and the concentrations of the species. Its explicit form is given in the Appendix.

The Navier-Stokes equation for the velocity resulting after substituting Eq. (\ref{3.1}) into Eq. (\ref{2.6}) will be now  linearized around the HCS. Firstly, the hydrodynamic fields are written in terms of their deviations from the HCS values,
\begin{equation}
\label{3.3}
n_{i}({\bm r},t) = n_{i,h}+ \delta n_{i}({\bm r},t),
\end{equation}
\begin{equation}
\label{3.4}
{\bm u} ({\bm r},t) = \delta {\bm u}({\bm r},t),
\end{equation}
\begin{equation}
\label{3.5}
T({\bm r},t) = T_{h}(t) + \delta T ({\bm r},t).
\end{equation}
Keeping  only up to first order in the deviations of the fields, the Navier-Stokes equation for the velocity becomes
\begin{equation}
\label{3.6}
\rho_{h}\, \frac{\partial \delta {\bm u}}{\partial t}+ n_{h} {\bm \nabla} \delta T + T_{h} {\bm \nabla} \delta n - \eta_{h} \left[ \nabla^{2} \delta {\bm u}+ \frac{d-2}{d} {\bm \nabla} \left( {\bm \nabla} \cdot \delta {\bm u} \right) \right]=0,
\end{equation}
with $\rho_{h}$ and $\eta_{h}$ being the mass density and the shear viscosity of the reference HCS, respectively.

At this point, it is convenient to introduce dimensionless time and space coordinates such that the time dependence of the reference state be eliminated in Eq. (\ref{2.6}). Then, new variables are defined by
\begin{equation}
\label{3.7}
{\bm l} \equiv \frac{\bm r}{\lambda_{h}}, \quad \tau \equiv \int_{0}^{t} dt^{\prime} \frac{v_{0}(T_{h})}{\lambda_{h}}\, .
\end{equation}
In the time scale $\tau$, the evolution equation for the temperature of the HCS becomes
\begin{equation}
\label{3.8}
\partial_{\tau} T_{h}(\tau)= - \zeta^{*} T_{h}( \tau),
\end{equation}
with the time-independent reduced cooling rate $\zeta^{*}$ given by
\begin{equation}
\label{3.9}
\zeta^{*} \equiv \frac{\lambda_{h} \zeta_{h}}{v_{0}(T_{h})}\, .
\end{equation}
Therefore, the temperature of the HCS decays exponentially on the $\tau$ scale.

Also, reduced hydrodynamic fields are introduced,
\begin{equation}
\label{3.10}
\varphi \equiv \frac{\delta n}{n_{h}}\, , \quad {\bm \omega} \equiv \frac{\delta {\bm u}}{v_{0}(T_{h})}\, , \quad \theta \equiv \frac{\delta T}{T_{h}}\, .
\end{equation}
 Use of the definitions in Eqs.\ (\ref{3.7}) and ({\ref{3.10}) into Eq.\ (\ref{3.6}) yields
 \begin{equation}
 \label{3.11}
 \left( \frac{\partial}{\partial \tau} - \frac{\zeta^{*}}{2} \right) {\bm \omega} + \beta \frac{\partial}{\partial {\bm l}} \left( \theta + \varphi \right) - \beta \eta^{*} \left[ \left( \frac{\partial}{\partial {\bm l}} \right)^{2} {\bm \omega} + \frac{d-2}{d} \frac{\partial}{\partial {\bm l}} \left( \frac{\partial {\bm \omega}}{\partial {\bm l}} \right) \right]=0,
 \end{equation}
with
\begin{equation}
\label{3.12}
\beta \equiv  \frac{\mu}{2(x_{1}m_{1}+x_{2}m_{2})}\, .
\end{equation}
Now, the Fourier representation defined as
\begin{equation}
\label{3.13}
\widetilde{f}_{\bm k} = \int d{\bm l}\, e^{-i {\bm k} \cdot {\bm l}} f({\bm l}),
\end{equation}
 for an arbitrary function $f({\bm l})$, will be employed. The transformed of Eq.\ (\ref{3.11}) is
\begin{equation}
\label{3.14}
\left( \frac{\partial}{\partial \tau}-\frac{\zeta^{*}}{2} \right) \widetilde{\bm \omega}_{\bm k}+
 i  \beta {\bm k} \left( \widetilde{\theta}_{\bm k}+\widetilde{\varphi}_{\bm k}\right) +\beta \eta^{*}
 \left[ k^{2}  \widetilde{\bm \omega}_{\bm k}+\frac{d-2}{d} {\bm k} ({\bm k}\cdot \widetilde{\bm \omega}_{\bm k} )\right]=0\, .
 \end{equation}
 Consider the transversal component $\widetilde{{\bm \omega}}_{\bm k, \perp}$  of the velocity field, i.e. the vector component of $\widetilde{\bm \omega}_{\bm k}$ perpendicular to ${\bm k}$. Its evolution equation is trivially obtained from Eq. (\ref{3.14}) and has the form
 \begin{equation}
 \label{3.15}
  \frac{\partial \widetilde{\bm \omega}_{{\bm k},\perp}}{\partial \tau}+
 \left(\beta \eta^{*} k^{2}-\frac{\zeta^{*}}{2}\right) \widetilde{\bm \omega}_{{\bm k},\perp}=0.
\, .
\end{equation}
This is a closed equation for $\widetilde{\bm \omega}_{{\bm k},\perp}$, similar to that found in one-component granular gases \cite{BDKyS98}. An equivalent equation has been derived in \cite{GMyD06}, where the general issue of the linear stability analysis of the Navier-Stokes equations of a dilute granular gas is addressed. The solution of Eq.\ (\ref{3.15}) reads
\begin{equation}
\label{3.16}
\widetilde{\bm \omega}_{{\bm k},\perp}(\tau)=e^{-s_{\perp}\tau} \widetilde{\bm \omega}_{{\bm k},\perp}(0)
\end{equation}
where the decay rate $s_{\perp}$ is
\begin{equation}
\label{3.17}
s_{\perp} \equiv \beta \eta^{*} k^{2}-\frac{\zeta^{*}}{2}\, .
\end{equation}
This leads to the identification of a critical value of the wavenumber vector given by
\begin{equation}
\label{3.18}
k_{c}= \left( \frac{\zeta^{*}}{2\beta \eta^{*}} \right)^{1/2}\, .
\end{equation}
A linear excitation of the scaled transversal velocity with $k<k_{c}$ grows in time. Therefore, vortices of the scaled velocity field are expected to develop in time when excitations of this kind are present in the system. This does not mean that the actual velocity field $\bm u$ is linearly unstable. In fact, it is easily seen that the perturbation $\delta {\bm u}$ decays exponentially in time because of Eq.\ (\ref{3.8}). The result above indicates that the linear analysis will eventually fail  and  nonlinear effects associated to coupling of hydrodynamic modes will have to be taken into account. In simple granular gases, this coupling leads to the development of the clustering instability. For this reason, the formation of vortices in the scaled velocity field is sometimes referred to as the shearing instability of the HCS.

\section{Mapping the HCS onto a steady state}
\label{s4}
In order to verify the validity of the ideas developed above and, in particular, to check whether the shear instability also exists in granular mixtures and if it is accurately predicted by the hydrodynamic Navier-Stokes equations,
the DSMC method \cite{Bi94,Ga00,BRyC96} has been used to generate numerical solutions of the coupled pair of Boltzmann equations. Actually, the method is an $N$-particle algorithm designed to mimic the dynamics of a low density gas and, therefore, it also provides equilibrium and non-equilibrium fluctuations and correlations.

One of the technical advantages of the DSMC method is that it permits to incorporate in the simulations the symmetries of the particular situation of interest. This allows a significant increase in the statistical accuracy of the measured properties. Here the aim is to investigate the development of inhomogeneities in the vicinity of the critical size associated with the shearing instability. Therefore, the simulation must allow the formation of spontaneous fluctuations of a given wavelength. For this reason, it is enough to consider a system in which gradients can occur in only one direction, arbitrarily taken as the $x$  axis. The components of the position of the particles perpendicular to that axis are not relevant from the point of view of the simulation. In other words, the simulation is restricted to systems which are  homogeneous in the planes perpendicular to the $x$ axis. The system size in the $x$ direction is $L$, and periodic boundary conditions are used in that direction.

A  simulation of the cooling mixture in the actual phase space variables is difficult since the rapid cooling of the system leads to rather small energies, and large uncertainties very soon. To deal with this, the procedure introduced in \cite{Lu01,BRyM04} for one-component granular gases and extended to mixtures in \cite{RyB13a}, will be employed. The idea is to exploit the existence of  an exact mapping of the HCS onto a steady state. Although the method can be formulated in the time scale $\tau$ defined in Eq. (\ref{3.7}), this would have the technical complication that the exact cooling rate is not known {\em a priori}. Consequently, it is convenient to introduce a new time scale $s$ by
\begin{equation}
\label{4.1}
ds = \frac{\zeta^{*}}{2 \varpi}\, d \tau = \frac{ \zeta^{*} v_{0}(T_{h})}{2 \varpi \lambda_{h}}\, dt,
\end{equation}
where $\varpi$ is an arbitrary dimensionless frequency. Now, the positions and velocities of the particles are represented in the ${\bm l}$ and $s$ scales. The particle dynamics in these variables consists of an accelerating streaming between collisions,
\begin{equation}
\label{4.2}
\frac{\partial {\bm l}}{\partial s} = {\bm \upsilon},
\end{equation}
\begin{equation}
\label{4.3}
\frac{\partial {\bm \upsilon}}{\partial s} = \varpi {\bm \upsilon},
\end{equation}
while the effect of the collision of two particles is the same as in the original time scale, given its instantaneous character.  The dynamics defined by Eqs. (\ref{4.2}) and (\ref{4.3})  is seen to be equivalent to a change in the original time scale,
\begin{equation}
\label{4.4}
 \varpi s = \ln \frac{t}{t_{0}},
 \end{equation}
where $t_{0}$ is another arbitrary constant. The acceleration term in the dynamics (\ref{4.3}) is able to balance the energy lost in collisions, thus enabling a steady state. The steady partial temperatures in the new dynamics are given by \cite{RyB13a}
\begin{equation}
\label{4.5}
T^{*}_{i,s}= \left( \frac{2 \varpi}{\overline{\zeta}_{i}} \right)^{2}, \quad \overline{\zeta}_{i} \equiv \frac{\zeta_{i}(T)}{T^{1/2}}\, .
\end{equation}
From Eqs. (\ref{2.13}) and (\ref{2.14}), it follows that also
\begin{equation}
\label{4.5a}
T^{*}_{s} = \left( \frac{2 \varpi}{\overline{\zeta}} \right)^{2},
\end{equation}
$\overline{\zeta} \equiv \zeta (T)/ T^{1/2}$.
The above mapping does not affect the hydrodynamic shear instability.  In the time scale $s$, Eq. (\ref{3.15}) becomes
\begin{equation}
\label{4.6}
\frac{ \partial \widetilde{\bm \omega}_{{\bm k} \perp}}{\partial s} + \left( \frac{2 \varpi \beta \eta^{*}k^{2}}{\zeta^{*}}- \varpi \right) \widetilde{\bm \omega}_{{\bm k} \perp}=0.
\end{equation}
As expected from dimensional analysis, the arbitrary constant $\varpi$ plays no role in the stability criterion.

Because of the symmetry of our simulations as described above, it is clear that the minimum wavevector allowed is given by $k_{min} =2 \pi \lambda_{h}/L$. Therefore, the stability condition $k_{min} > k_{c}$ with $k_{c}$ given by Eq. (\ref{3.18}) is equivalent to $L <L_{c}$ with the critical length $L_{c}$ given by
\begin{equation}
\label{4.7}
L_{c}= 2\pi \lambda_{h} \left( \frac{2 \beta \eta^{*}}{\zeta^{*}} \right)^{1/2}\, .
\end{equation}
It is worth to mention the existence of a related instability for the total momentum of the system in the scaled variables \cite{Lu01}. Nevertheless, it is not physically relevant and can be eliminated by taking a vanishing initial  total momentum.

\section{Simulation results}
\label{s5}
In the simulations to be reported in the following, a binary mixture of $N=N_{1}+N_{2}$ inelastic hard spheres ($d=3$) has been used. In order to reduce the number of parameters characterizing the system and to allow for a systematic study of those being varied, the number of particles of both species, and therefore the concentrations, have always been the same ($N_{1}=N_{2}$), as well as the diameters of the particles, i.e. $\sigma_{1}=\sigma_{2}$. Moreover, the coefficient of normal restitution for collisions between particles of different species has been taken as the average of the coefficients for equal species collisions, i.e. $\alpha_{12} = (\alpha_{11}+\alpha_{22})/2$. On the other hand,  the mass ratio $\Delta \equiv m_{2}/m_{1}$ and the two coefficients of normal restitution $\alpha_{11}$ and $\alpha_{22}$, as well as the size $L$ of the system have been varied in the simulations.

The behaviour of several properties of the HCS as the size $L$ of the system approaches the critical value has been studied. The number of particles per unit of length in the $x$ direction has been kept fixed in the simulations, $N_{x} \equiv N/L_{x}= 2000 \ell_{h}^{-1}$, where $\ell_{h} =  \lambda_{h} / \sqrt{2} \pi = \left(\sqrt{2} \pi \sigma^{2} n_{h} \right)^{-1}$ is the mean free path. It is worth to stress that the number of particles used in the DSMC method does not affect the validity of the low density limit, that is inherent to the method itself \cite{Bi94}.

The simulations were performed using the steady representation of the HCS discussed in the previous section and, unless explicitly otherwise established, the values of the properties reported in the following have been averaged in time once the system had reached the steady state, as well as over a number of different trajectories (typically 50). The value of the arbitrary constant $\varpi$ was chosen in all cases as $\varpi = \overline{\zeta}_{G}/2$, with $\overline{\zeta}_{G}$ being the value of $\zeta$ obtained in the Gaussian approximation, i.e. those given by Eq.\ (\ref{2.15}). If the Gaussian approximation were exact, the measured steady value of the total temperature would have been one.

The first point addressed in the simulations was to check that the scaled transversal velocity field was really the first hydrodynamic mode becoming unstable as $L$ increases. The steady state reached by the system for different sizes was investigated, starting from a system with $L$ small enough as to guarantee that the HCS was stable, and increasing $L$ from there on. The different hydrodynamic fields were monitored at different times, and in all cases it was found that they were the $y$ and $z$ components of the scaled velocity field the first hydrodynamic modes exhibiting large fluctuations.  As long as $L$ is not large, these fluctuations eventually decay, but when
the system size was increased enough, a non-decaying scaled transversal velocity field emerged. This is illustrated in Fig.\ \ref{fig1}, where the density and one of the components of the transversal velocity field are shown at three different times for a system with $\alpha_{11}=0.8$, $\alpha_{22}=0.98$, $m_{2}/m_{1}=4$, and $L=23.7 \ell_{h}$. Both the total hydrodynamic fields as well as those associated with each of the species are displayed. The velocity field for each of the components are defined by equations similar to Eq. (\ref{2.2}) \cite{DyB11}, and they have been scaled with the square root of the temperature of the system. It is seen that the system exhibits an spontaneous perturbation of the transversal velocity field that does not decay in time, and corresponds to the first possible harmonic. On the other hand, the density remains homogeneous. Note that the local average velocities of the species is the same as that of the whole fluid. Then, it was concluded that the scaled shear mode is the field for which the linear approximation first breaks down.

\begin{figure}
\includegraphics[scale=0.5,angle=0]{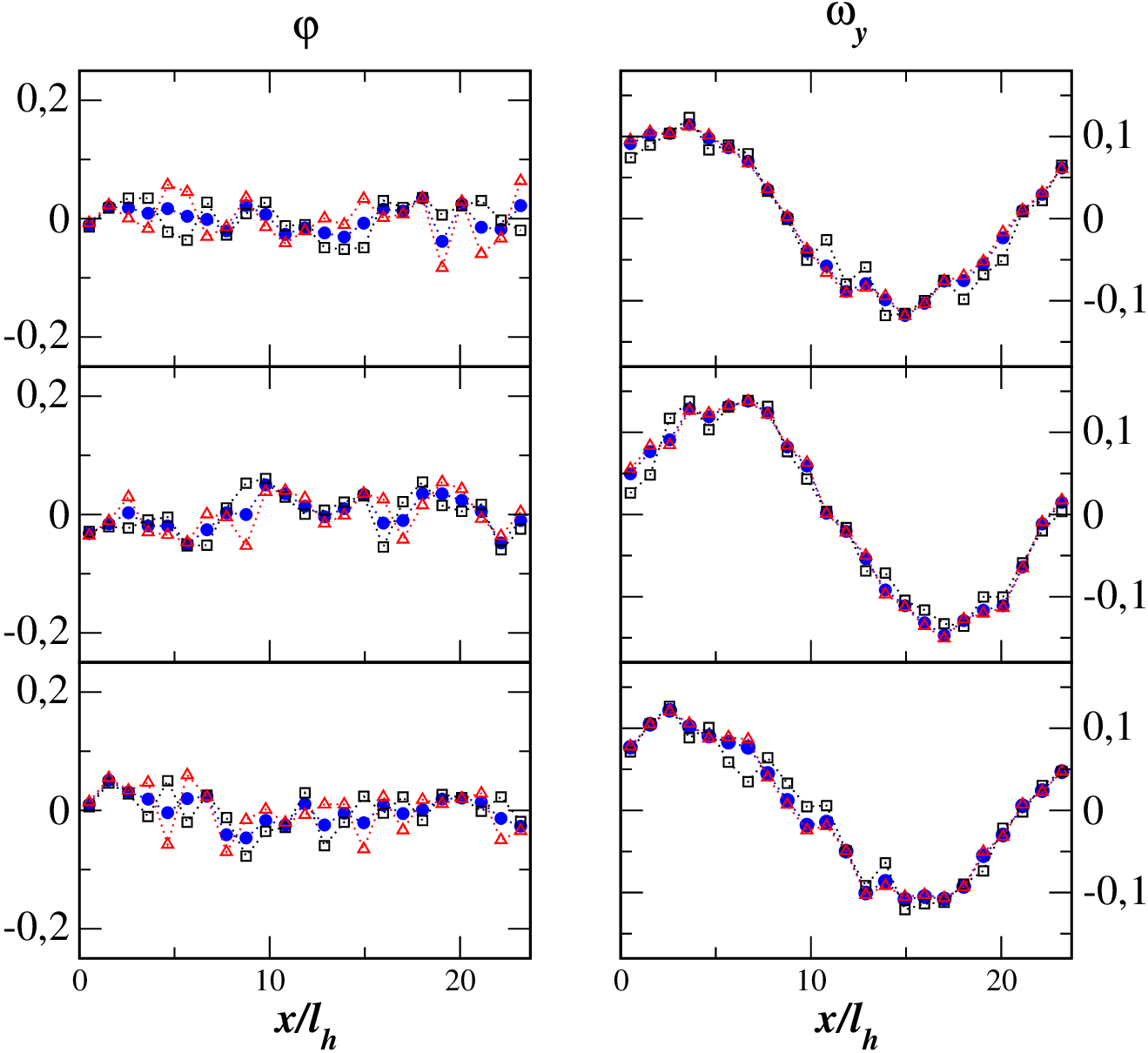}
\caption{(Color online) Snapshots of the dimensionless scaled density (left) and of the one of the components of the perpendicular velocity field $\omega_{y}$ for a system with $\alpha_{11}=0.8$, $\alpha_{22}=0.98$, $m_{2}/m_{1}=4$, and $L= 23.7 \ell_{h}$. The (blue) circles correspond to the whole fluid properties, while the empty (black) square and (red) triangles are for species $1$ and $2$, respectively. From top to bottom the times are $s=6.68 \times 10^{3}$, $1.33 \times 10^{4}$, and $ 1.99 \times 10^{4}$, in the dimensionless scale defined by Eq. (\protect{\ref{4.1}}), with $\varpi$ chosen as discussed in the main text.}  \label{fig1}
\end{figure}

To measure the critical size $L_{c}$, the following procedure was used. First, the average value of the total energy in the steady state $E$ was measured as a function of the size of the system. Then, it was assumed, to be checked in the simulation results, that the behaviour of the average steady energy near but below the shearing instability obeys a law of the form
\begin{equation}
\label{5.1}
\delta_{E} \equiv \frac{ <E> - <E>_{h}}{<E>_{h}} \propto \left( \frac{L_{c}-L}{L_{c}} \right)^{-1} \equiv \delta_{L}^{-1}\, ,
\end{equation}
where $<f>_{h}$ denotes the constant asymptotic average value of the property $f$ in the HCS, far away from the shear instability, also obtained from the simulations. The above behaviour is suggested by the results obtained for a one-component dilute granular gas near its shear instability \cite{BGMyR05,ByR07}.

In Fig. \ref{fig2}, $\delta_{E}^{-1}$ is plotted as a function of the system size. The observed linear behaviour is consistent with Eq.\ (\ref{5.1}), and from the parameters of the linear fits, simulation values of the critical size $L_{c}$ are directly obtained. The fits of the three lines lead to the same value, namely $L_{c} \simeq 30.13 \ell_{h}$. A similar behaviour was obtained for all the values of the restitution coefficients and the mass ratio investigated. It follows that the increase of the total average energy of the system and also that of each of the components, as the system approaches the instability, is characterized by Eq.\ (\ref{5.1}).

\begin{figure}
\includegraphics[scale=0.5,angle=0]{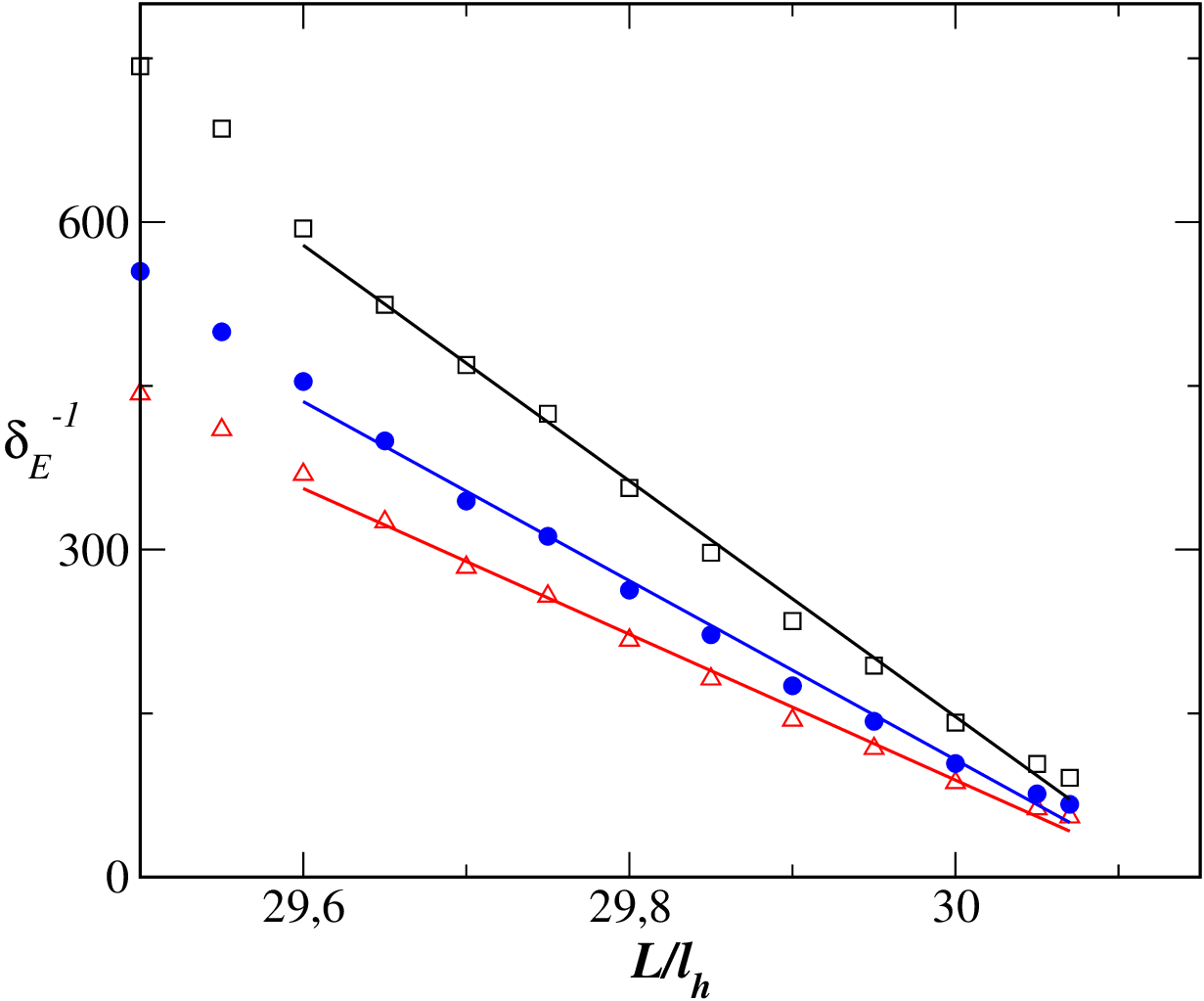}
\caption{(Color online) Relative deviations $\delta_{E} \equiv (<E>-<E>_{h})/<E>_{h}$  of the average total energy of the system from its asymptotic values in the HCS, as a function of the system size $L$, in the vicinity of the shear instability. The parameters of the system are $\alpha_{11}=0.92$, $\alpha_{22}=0.98$, and $m_{2}/m_{1} = 8$. The symbols are from the simulations and the straight lines are fits in the ``critical region''. The (blue) circles correspond to the whole fluid, while the empty (black) squares and (red) triangles are for species $1$ and $2$, respectively.}  \label{fig2}
\end{figure}

The comparison between the measured critical sizes and the theoretical prediction given by Eq. (\ref{4.7}) is presented in Fig.\ \ref{fig3} as a function of the mass ratio $m_{2}/m_{1}$, for three different sets of values of the coefficients of normal restitution. The agreement is quite good over the two decades considered. The non-monotonic dependence of the critical length on the mass ratio for given coefficients of normal restitution must be noticed. This is specially relevant for strong inelasticities.

\begin{figure}
\begin{tabular}{c}
\includegraphics[scale=0.4,angle=0]{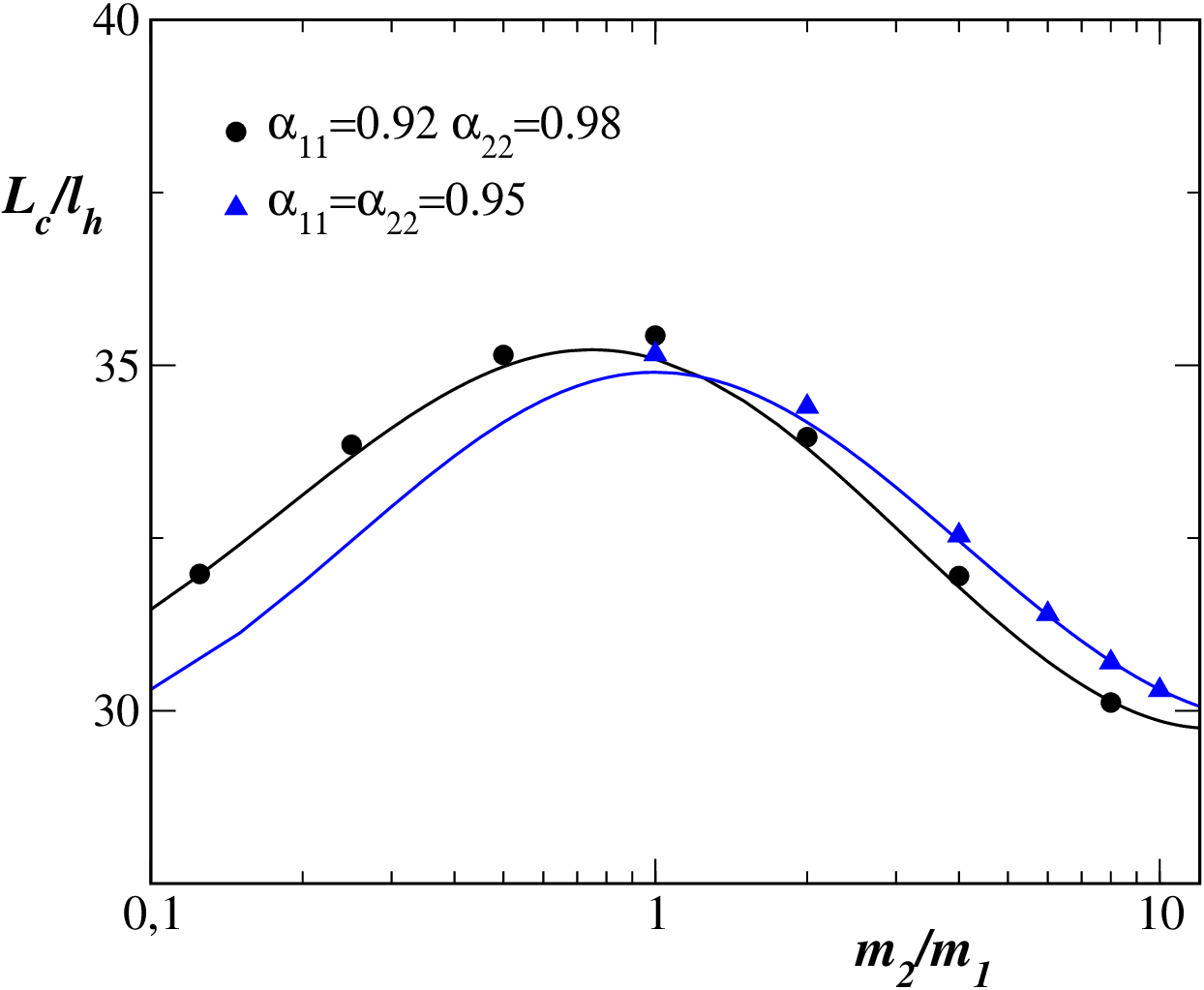}
\\
\vspace{1cm}
\\
\includegraphics[scale=0.4,angle=0]{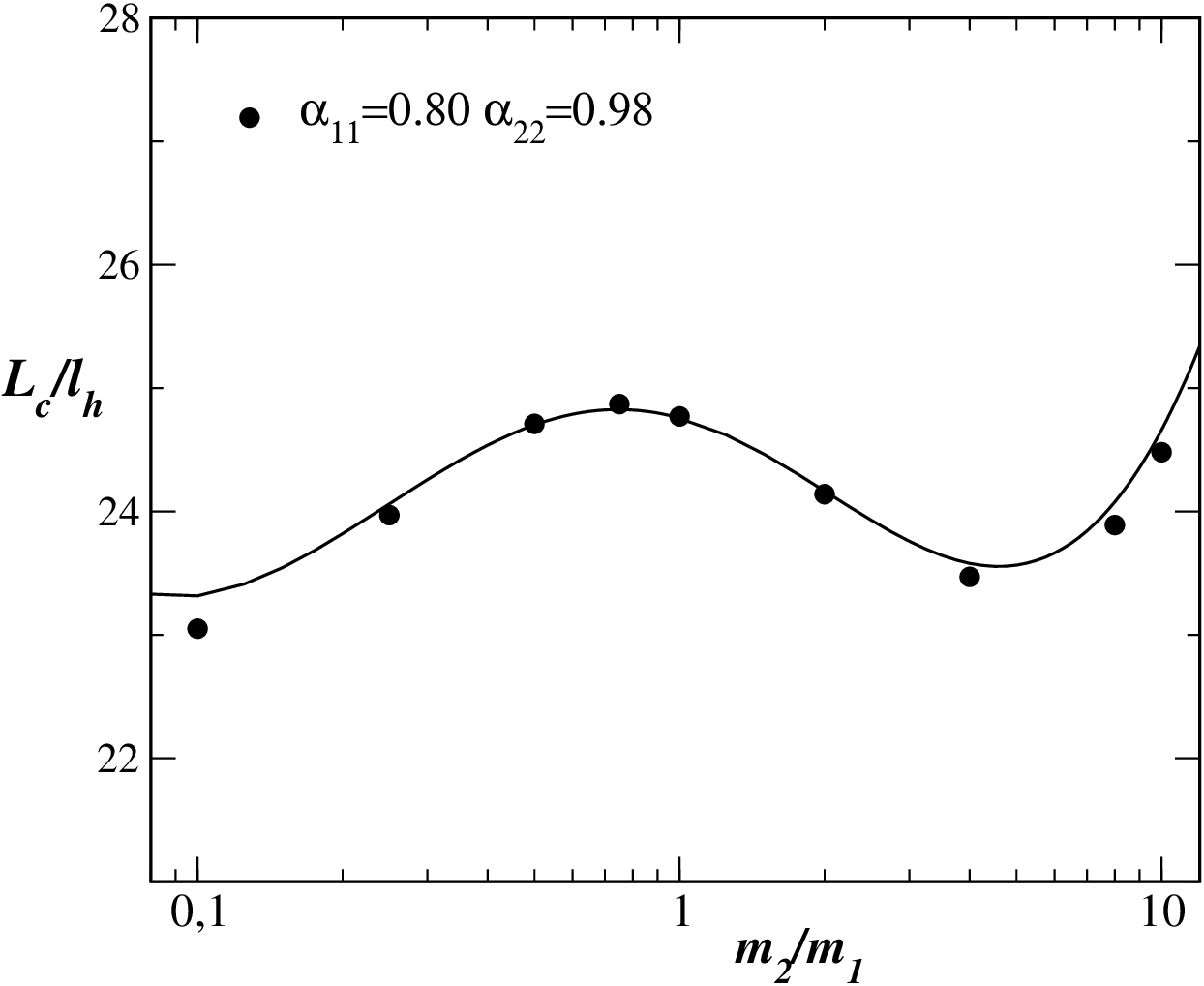}
\end{tabular}
\caption{(Color online) Dimensionless critical size $L_{c}/ \ell_{h}$ for the shear instability of the HCS as a function of the mass ratio $m_{2}/m_{1}$. The symbols are from the DSMC method, while the solid lines are the theoretical predictions given by Eq.\ (\protect{\ref{4.7}}). Three different sets of values of the coefficients of normal restitution have been considered, as indicated in the insets.}  \label{fig3}
\end{figure}

Another quantity investigated in the simulations is the temperature ratio, $\gamma_{21} \equiv T_{2}/T_{1}$. Notice that as a consequence of Eqs. (\ref{2.14}) and (\ref{4.5}), it is $T_{2,h}(t)/T_{1,h}(t) = T_{2,s}^{*}/T_{1,s}^{*}$. For mixtures whose components have very dissimilar masses, a small but systematic deviation of $\gamma_{21}$ from its HCS value, $\gamma_{21,h}$, was observed when the system approaches its critical size. This deviation is larger the closer the length of the system to its critical value. It is found that $\gamma_{21} > \gamma_{21,h}$ for $m_{2} >m_{1}$, while $\gamma_{21} < \gamma_{21,h}$ for $m_{2}  <m_{1}$. Finally, for equal masses of both components, no deviation from the HCS value is observed. In any case, it must be noticed that the deviations from the HCS values are never larger than $1 \%$.  This behaviour is shown in Fig. \ref{fig4}, in which $\gamma_{21}$ is plotted as a function of the reduced distance to the critical point $\delta_{L} \equiv (L_{c}-L)/L_{c}$ for a system with $\alpha_{11}=0.92$ and $\alpha_{22}=0.98$. Results for several values of the mass ratio, $\Delta \equiv m_{2}/m_{1}$ are displayed, as indicated in the figure caption.

\begin{figure}
\includegraphics[scale=0.5,angle=0]{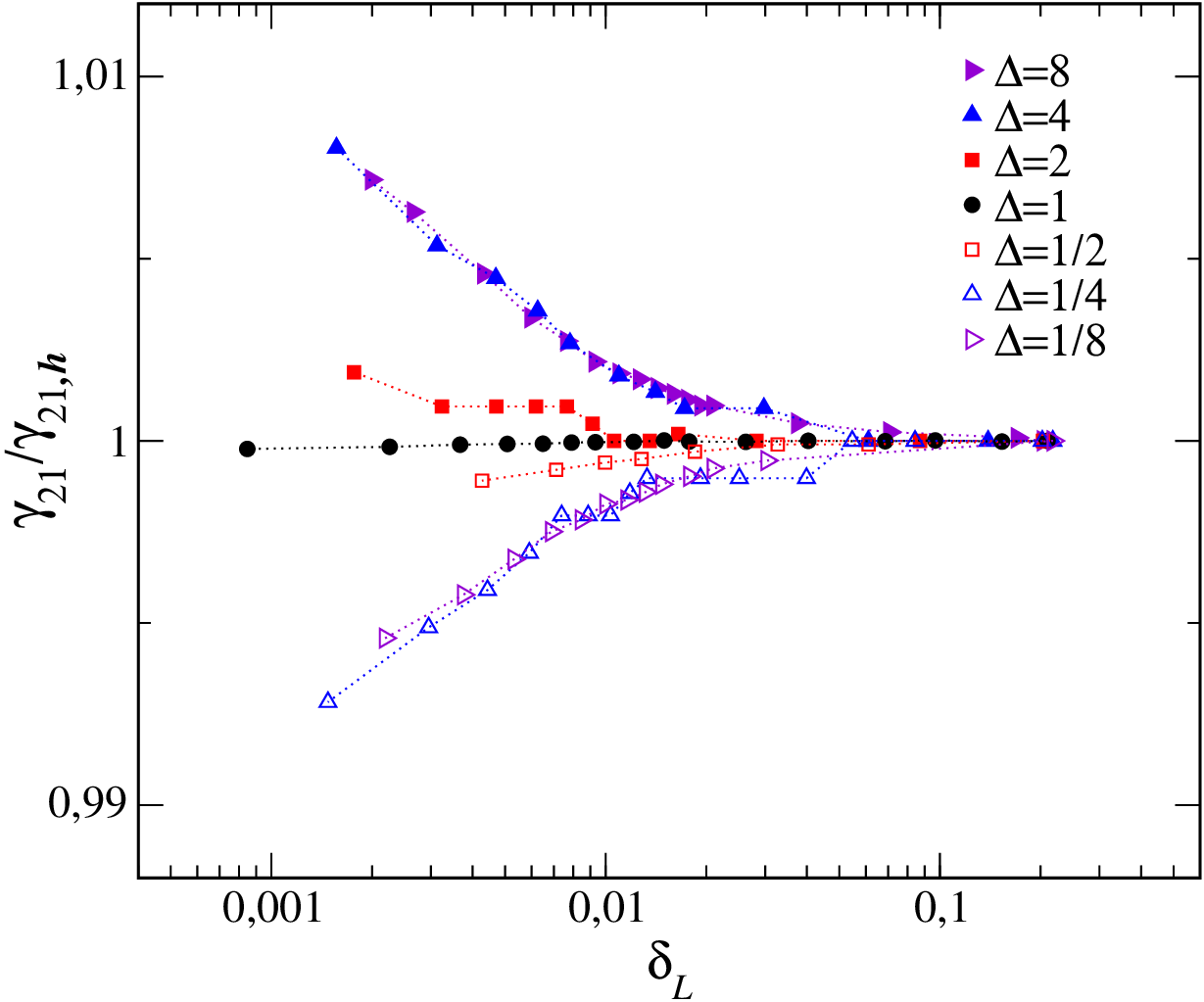}
\caption{(Color online) Temperature ratio $\gamma_{21} \equiv T_{2}/T_{1}$ in the steady state divided by its HCS value $\gamma_{21,h}$, as a function of the relative ``distance'' to the critical size, $\delta_{L}$. The coefficients of normal restitution are $\alpha_{11}=0.92$ and $\alpha_{22}=0.98$. Results for different mass ratios $\Delta \equiv m_{2}/m_{1}$ are shown, as indicated in the inset.}  \label{fig4}
\end{figure}

\section{Final comments}
\label{s6}
The results presented in this paper show that a freely evolving dilute granular mixture exhibits an instability associated to the transversal shear modes that is similar to the one occurring in one-component granular gases. The existence of the instability, and the parameters characterizing the critical point, are accurately predicted by the linearized hydrodynamic equations to Navier-Stokes order. Although the detailed nonlinear mechanisms leading to the formation of density clusters beyond the shear  instability have not been investigated here, it seems evident that they are the same as those for one-component systems \cite{BRyC99}, given the similarity of the behavior of both, mixtures and simple systems, when approaching the critical size \cite{ByR03}.

Another relevant quantity to characterize the system near the instability is the second moment of the fluctuations of the total energy, defined as
\begin{equation}
\label{6.1}
\Sigma^{2} \equiv \frac{N (<E^{2}>-<E>^{2})}{<E>^{2}}.
\end{equation}
The factor $N$ has been introduced to scale out the dependence due to the number of particles (or size $L$) of the system \cite{BGMyR05}. Consider
\begin{equation}
\label{6.2}
\delta_{\Sigma^{2}} \equiv \frac{ \Sigma^{2} - \Sigma_{h}^{2}}{\Sigma_{h}^{2}},
\end{equation}
where $\Sigma_{h}$ is the steady value of $\Sigma$ far away from the instability. In one-component gases, it was found that near but below the shearing instability,
\begin{equation}
\label{6.3}
\delta_{\Sigma^{2}} \propto \delta_{L}^{-2},
\end{equation}
with $\delta_{L}$ given by Eq.\ (\ref{5.1}). Our simulation results clearly indicate that this relation also holds for mixtures. Actually, if the numerical results for the energy dispersion  are fitted to it, and the fitting parameters are used to determine the critical length $L_{c}$, the values are the same, within the statistical errors, as those obtained from the critical behaviour of the average energy and  discussed above.

To analyze in more detail the nature of the energy fluctuations, its distribution function was measured in the simulations. Again prompted by the properties of the critical region in one-dimensional granular gases
\cite{BGMyR05}, the quantity
\begin{equation}
\label{6.3a}
\epsilon \equiv \frac{E- <E>}{<(E-<E>)^{2} >^{1/2}}
\end{equation}
is considered. Far away from the instability, i.e. $L \ll L_{c}$, the probability distribution of $\epsilon$ is Gaussian, as expected. Nevertheless, as the instability is approached, the distribution strongly deviates from a Gaussian, showing a clear asymmetry around the mean value.  Moreover, and quite surprisingly, close enough to the critical length, the data for different values of the restitution coefficients, of the mass ratio, and of  the length of the system, collapse onto the same curve, as illustrated in Fig.\ \ref{fig5}. This indicates that all the dependence of the distribution on the parameters of the system occurs through the second moment of the distribution. In addition, the shape of the distribution is the same as that found in one-component granular gases \cite{BGMyR05}, as well as in other equilibrium and non-equilibrium systems \cite{BHyP98,Betal00}. In these cases, an accurate expression to fit the data is
\begin{equation}
\label{6.4}
P_{0}(y)= K \left(e^{x-e^{x}} \right)^{\pi /2}, \quad x=b(y-s).
\end{equation}
The values of the parameters in the above distribution follow from the normalization, zero mean, and unit variance conditions, with the result $K=2.14$, $b=0.938$, and $s=0.374$. Therefore, it has no fitting parameters. In Fig.\ \ref{fig5} the solid line is the plot of $P_{0} (-\epsilon)$. A remarkable agreement with the simulation data is obtained. A peculiarity of the present case as compared with the other systems in which the  distribution (\ref{6.4}) has been used, is that here it is the symmetric with respect to the origin the one fitting the numerical data. While in granular gases, large positive fluctuations are more frequent than their symmetric, it happens the other way around in the molecular systems considered in \cite{BHyP98,Betal00}. It is possible that this difference be due to the dissipative character of granular systems.

It is worth mentioning  that in the case of one-component granular gases, the critical behaviour of the system near the shearing instability can be, at least qualitatively, understood in terms of nonlinear fluctuating hydrodynamics couplings  \cite{BDGyM06}. It is expected that the analysis can be extended to binary mixtures with the same degree of accuracy.

\begin{figure}
\includegraphics[scale=0.5,angle=0]{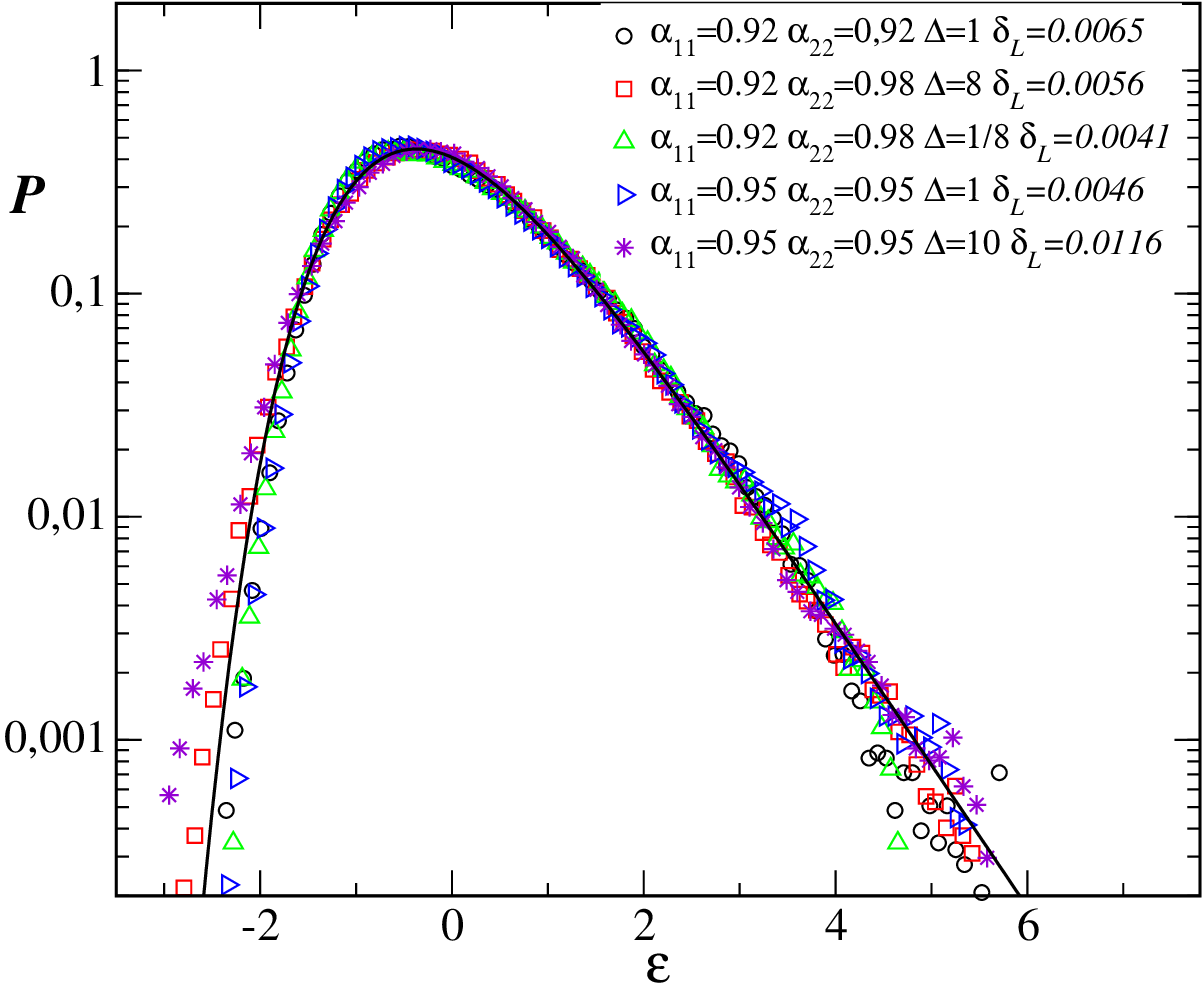}
\caption{(Color online) Probability density function of the scaled relative energy fluctuations $\epsilon$ close to the critical size $L_{c}$. Results for several values of the coefficients of normal restitution $\alpha_{ij}$, the mass ratio $\Delta \equiv m_{2}/m_{1}$, and the scaled length $\delta_{L} \equiv (L-L_{c})/L_{c}$ are shown, as indicated in the inset. The symbols are from simulations, and the solid line is the distribution function given in Eq. (\protect{\ref{6.4}}), changing $y$ into $-\epsilon$. }  \label{fig5}
\end{figure}

\section{Acknowledgements}

This research was supported by the Ministerio de Educaci\'{o}n y Ciencia (Spain)
through Grant No. FIS2011-24460 (partially financed by FEDER funds). We want to thank V. Garz\'{o} for providing us with the corrected expression for the shear viscosity coefficient given in the Appendix.

\appendix
 \section{Reduced viscosity of a dilute binary mixture}
 \label{secap:1}
 The expression for the reduced viscosity $\eta^{*}$ of a binary mixture of inelastic hard particles has been obtained in
 \cite{GyM07}. It is given here for the sake of completeness and also because there is an error in the results reported in the aforementioned
 reference.
 The expression for the reduced viscosity reads
 \begin{equation}
 \label{b.1}
 \eta^{*}=x_{1}\gamma_{1}^{2} \eta_{1}^{*}+x_{2}\gamma_{2}^{2} \eta_{2}^{*},
 \end{equation}
 with
 \begin{equation}
 \label{b.2}
 \eta_{1}^{*}=\frac{2}{\gamma_{1}\gamma_{2}} \, \frac{\gamma_{2} (2\tau_{22}-\zeta^{*})-2 \gamma_{1} \tau_{12}}
 {\zeta^{*\,2}-2\zeta^{*} (\tau_{11}+\tau_{22})+4 (\tau_{11} \tau_{22}-\tau_{12}\tau_{21})}\, .
 \end{equation}
 The coefficients $\tau_{11}$ and $\tau_{12}$ are given by
 \[
  \tau_{11}=\frac{2\pi^{(d-1)/2}}{d (d+2) \Gamma\left(\frac{d}{2}\right)}
  \bigg\{ \left( \frac{\sigma_{1}}{\sigma_{12}} \right)^{1/2} x_{1} (2\theta_{1})^{-1/2}
 (3+2 d-3 \alpha_{11})(1+\alpha_{11}) \]
 \[ + 2 x_{2}\mu_{21} (1+\alpha_{12}) \theta_{1}^{3/2} \theta_{2}^{-1/2}
 \Big[ (d+3) (\mu_{12} \theta_{2} -\mu_{21}\theta_{1}) \theta_{1}^{-2} (\theta_{1}+\theta_{2})^{-1/2}\]
 \begin{equation}
 \label{b.3}
% \left.\left.
 +\frac{3+2d-3\alpha_{12}}{2} \mu_{21} \theta_{1}^{-2} (\theta_{1}+\theta_{2})^{1/2} +
 \frac{ 2d(d+1)-4}{2 (d-1)} \theta_{1}^{-1} (\theta_{1}+\theta_{2})^{-1/2}\Big] \bigg\} \, ,
 \end{equation}
 \[\tau_{12}=\frac{4\pi^{(d-1)/2}}{d (d+2) \Gamma\left(\frac{d}{2}\right)} x_{2} \frac{\mu_{21}^{2}}{\mu_{12}}
 \theta_{1}^{3/2} \theta_{2}^{-1/2} (1+\alpha_{12})\bigg[ (d+3)(\mu_{12} \theta_{2}-\mu_{21} \theta_{1})
 \theta_{2}^{-2} (\theta_{1}+\theta_{2})^{-1/2}  \]
 \begin{equation}
 \label{b.4}
 +\frac{3+2 d-3\alpha_{12}}{2} \mu_{21} \theta_{2}^{-2} (\theta_{1}+\theta_{2})^{1/2}-
 \frac{2 d (d+1)-4}{2 (d-1)} \theta_{2}^{-1} (\theta_{1}+\theta_{2})^{-1/2} \bigg]\, .
 \end{equation}
 The quantities $\mu_{ij}$, $\theta_{i}$, and $\sigma_{12}$ have been defined in Eqs. (\ref{2.18}), (\ref{2.19}), and above Eq.\ (\ref{2.16}), respectively. Moreover,
 \begin{equation}
 \label{b.5}
 \sigma_{12} \equiv \frac{\sigma_{1}+ \sigma_{2}}{2}\, .
 \end{equation}
The expression for the reduced contribution viscosity $\eta_{2}^{*}$ can bo obtained from Eqs. (\ref{b.1})-(\ref{b.5}) just
 interchanging the indexes 1 and 2.


\begin{thebibliography}{99}

\bibitem{Go03} I. Goldhirsch, {\em Rapid Granular Flows}, Ann. Rev. Fluid Mech. {\bf 35}, 267 (2003).

\bibitem{BDyS97} J.J. Brey, J.W. Dufty, and A. Santos, J. Stat. Phys. {\bf 87}, 1051 (1997).

\bibitem{GyZ93} I. Goldhirsch and G. Zanetti, Phys. Rev. Lett. {\bf 70}, 1619 (1993); I. Goldhirsch, M.L. Tan, and G. Zanetti, J. Sci. Comput. {\bf 8}, 1 (1993).

\bibitem{MnyY94} S. McNamara and W.R. Young, Phys. Rev. E {\bf 50}, R28 (1994).

\bibitem{ByP04} N.V. Brilliantov and T. P\"{o}schel, {\em Kinetic Theory of Granular Gases} (Oxford University Press, Oxford, 2004).

\bibitem{BRyC99} J.J. Brey, M.J. Ruiz-Montero, and D. Cubero, Phys. Rev. E {\bf 60}, 3150 (1999).

\bibitem{vNEByO97} T.P.C. van Noije, M.H. Ernst, R. Brito, and J.A.G. Orza, Phys. Rev. Lett. {\bf 79}, 411 (1997).

\bibitem{BMyR98}J.J. Brey, F. Moreno, and M.J. Ruiz-Montero, Phys. Fluids {\bf 10}, 2965 (1998).


\bibitem{BDGyM06} J.J. Brey, A. Dom\'{\i}nguez, M.I. Garc\'{\i}a de Soria, and P. Maynar, Phys. Rev. Lett. {\bf 96}, 158002 (2006).

\bibitem{GyM07} V. Garz\'{o} and J.M. Montanero, J. Stat. Phys. {\bf 129}, 27 (2007).

\bibitem{GHyD07} V. Garz\'{o}, C.M. Hrenya, and J.W. Dufty, Phys. Rev. E {\bf 76}, 031304 (2007).

\bibitem{GMyD06} V. Garz\'{o}, J.M. Montanero, and J.W. Dufty, Phys. Fluids {\bf 18}, 083305 (2006).


\bibitem{GyD99} V. Garz\'{o} and J.W. Dufty, Phys. Rev. E {\bf 60}, 5706 (1999).

\bibitem{DByL02} J.W. Dufty, J.J. Brey, and J. Lutsko, Phys. Rev. E {\bf 65}, 051303 (2002).

\bibitem{ByT02} A. Barrat and E. Trizac, Gran. Matter {\bf 4}, 57 (2002).



\bibitem{ByR11} J.J. Brey and M.J. Ruiz-Montero, Phys. Rev. E {\bf 84}, 031302 (2011).

\bibitem{DHGyD02} S.R. Dahl, C.M. Hrenya, V. Garz\'{o}, and J.W. Dufty, Phys. Rev. E {\bf 66}, 041301 (2002).

\bibitem{MyG02} J.M. Montanero and V. Garz\'{o}, Granular Matter {\bf 4}, 17 (2002).

\bibitem{RyB13a} M.J. Ruiz-Montero and J.J. Brey, AIP Conf. Proc. {\bf 1501}, 977 (2012).



\bibitem{BDKyS98} J.J. Brey, J.W. Dufty, C.S. Kim, and A. Santos, Phys. Rev. E {\bf 58}, 4638 (1998).




\bibitem{GyD02} V. Garz\'{o} and J.W. Dufty, Phys of Fluids {\bf 14}, 1476 (2002).

\bibitem{Bi94} G. Bird, {\em Molecular Gas Dynamics and the Direct Simulation of Gas Flows} (Clarendon, Oxford, 1994).

\bibitem{Ga00} A. Garc\'{\i}a, {\em Numerical methods for Physics} (Prentice Hall, Englewood Hills, NJ, 2000).

\bibitem{BRyC96} J.J. Brey, M.J. Ruiz-Montero, and D. Cubero, Phys. Rev. E {\bf 54}, 3664 (1996).

\bibitem{Lu01} J.F. Lutsko, Phys. Rev. E {\bf 63}, 061211 (2001).

\bibitem{BRyM04} J.J. Brey, M.J. Ruiz-Montero, and F. Moreno, Phys. Rev. E {\bf 69}, 051303 (2004).

\bibitem{DyB11} J.W. Dufty and J.J. Brey, Math. Model. Nat. Phenom. {\bf 6}, 19 (2011).



\bibitem{BGMyR05} J.J. Brey, M.I. Garc\'{\i}a de Soria, P. Maynar, and M.J. Ruiz-Montero, Phys. Rev. Lett {\bf 94}, 098001 (2005).

\bibitem{ByR07} J.J. Brey and M.J. Ruiz-Montero, Granular Matt. {\bf 10}, 53 (2007).

\bibitem{ByR03} J.J. Brey and M.J. Ruiz-Montero, unpublished.

\bibitem{BHyP98} S.T. Bramwell, P.C.W. Holdsworth, and J.-F. Pinton, Nature {\bf 396}, 552 (1998).

\bibitem{Betal00} S.T. Bramwell, K. Christensen, J.-Y. Fortin, P.C.W. Holdsworth, H.J. Jensen, S. Lise, J.M. L\'{o}pez, M. Nicodemi, J.-F. Pinton, and M. Sellitto, Phys. Rev. Lett. {\bf 84}, 3744 (2000).


\end{thebibliography}
\end{document}